# Accelerated Mapping of Electronic Density of States Patterns of Metallic Nanoparticles via Machine-Learning

Kihoon Bang,[†] Byung Chul Yeo,[‡] Donghun Kim,[‡] Sang Soo Han,[‡,]* Hyuck Mo Lee[†,]*

[†]Department of Materials Science and Engineering, Korea Advanced Institute of Science and Technology (KAIST), 291 Daehak-ro, Yuseong-gu, Daejeon 34141, Republic of Korea

[‡]Computational Science Research Center, Korea Institute of Science and Technology (KIST), 5 Hwarangno 14-gil, Seongbuk-gu, Seoul 02792, Republic of Korea

**Corresponding Author**

*E-mail: sangsoo@kist.re.kr (S.S.H.)

*E-mail: hmlee@kaist.ac.kr (H.M.L)

**Abstract**

Within first-principles density functional theory (DFT) frameworks, accurate but fast prediction of electronic structures of nanoparticles (NPs) remains challenging. Herein, we propose a machine-learning architecture to rapidly but reasonably predict electronic density of states (DOS) patterns of metallic NPs via a combination of principal component analysis (PCA) and the crystal graph convolutional neural network (CGCNN). By applying PCA, one can convert a mathematically high-dimensional DOS image to a low-dimensional vector. The CGCNN plays a key role in reflecting the effects of local atomic structures on the DOS patterns of NPs with only a few of material features (e.g., melting temperature, the number of $d$ electrons, and atomic radius) that are easily obtained from a periodic table. The PCA-CGCNN model is applicable for all pure and bimetallic NPs, in which a handful DOS training sets that are easily obtained with the typical DFT method, such as bulk, slab, and small-sized NPs, are considered. Although there is a small loss of accuracy with the PCA-CGCNN method compared to DFT calculations, the prediction speed is much faster than that of DFT methods and is not nearly as affected by the system sizes of NPs. Our approach not only can be immediately applied to predict electronic structures of actual nanometer scaled NPs to be experimentally synthesized, but also be used to explore correlations between atomic structures and other spectrum image data of the materials (e.g., X-ray diffraction, X-ray photoelectron spectroscopy, and Raman spectroscopy).

**INTRODUCTION**

Nanoparticles (NPs) are of great scientific interest because they often show unexpected physical and chemical properties resulting from their quantum confinement effect[1,2] or high surface area[3,4]. This leads to various applications of NPs, such as quantum dots,[5-7] magnetic[8,9] or bio-[10-13] materials, and catalysis[3,14-21]. As a key feature to determine the properties of NPs, an electronic structure such as electronic density of states (DOS) has been usually considered, where the electronic structure significantly depends on the sizes and shapes of the NPs although the elements constituting the NPs are identical.[9,17,20,22-26] First-principles density functional theory (DFT) calculations have been mainly utilized to predict DOS patterns of NP structures. In particular, the plane-wave (PW) basis technique has been employed for metallic NP systems despite its extremely high computational cost. Moreover, NP structures require a much higher computational cost than bulk or slab structures. In the case of the PW technique, it is necessary that the entire simulation box, including the vacuum space, must be filled with PWs, seriously reducing the computational speed.[27] In this regard, the fast but accurate electronic structure calculation for metallic NPs still remains challenging.

To bypass the first-principles framework, a machine-learning (ML) approach has been recently pursued.[28-32] In particular, Chandrasekaran et al.[29] developed a neural network (NN) model for the prediction of DOS patterns and showed that its computational cost was linearly scaled with system size ($N$) [O($N$)], while the DFT method was scaled as O($N^2$). With a similar aim, Yeo et al.[30] developed an ML scheme based on principal component analysis (PCA) and successfully applied it to bulk and slab structures of multicomponent metallic systems although it is a very simple process. Moreover, it showed a computational cost independent of the system size. Despite such success, the scheme predicts the DOS pattern of a test system via a linear interpolation between the two training systems that is most similar to the test composition, which likely reduces the versatility of the scheme. When mapping the DOS patterns of materials, it is important to appropriately reflect the local environments of each atom in the structures because the DOS patterns are sensitive to the local atomic environment.

Metallic NP structures can be regarded as consisting of core and shell regions. Here, although the core region can be treated as a bulk structure, the shell region is an assembly consisting of surface atoms with various coordination numbers, which motivates us to improve the previous PCA-based method by more elaborately learning the local environments of atoms in NPs when predicting their DOS patterns. Xie and Grossman[33] reported a crystal graph convolutional neural network (CGCNN) framework enabling a universal and interpretable representation of crystalline materials. This model converts atomic structures in bulks to graphs, and then the graph fingerprints learn the local environments of atoms by an additional CNN process. Recently, we also demonstrated that the framework can be extended to slab structures.[21] These facts reveal that the CGCNN is readily applicable for representing atomic structures of NPs, and a combination of PCA and the CGCNN is expected to provide a reasonable and fast mapping of DOS patterns of NPs.

In this work, we propose an ML paradigm to predict DOS patterns (both of shapes and of values) of metallic NPs through a combination of PCA and the CGCNN, where the model is learned with DOS patterns of bulk, slab, and small-sized NPs (e.g., $Pt_{13}$) that are not time-consuming to obtain with the state-of-the-art DFT calculations. Within the PCA-CGCNN framework, one can predict DOS patterns for not only pristine NPs but also alloyed ones with a small loss of accuracy compared to DFT calculations, where effects of the sizes and shapes of metallic NPs have also been explored. Moreover, the method shows a computational cost nearly independent of the system size.

**RESULTS AND DISCUSSION**

Because the mathematical dimension of a DOS pattern is very high (e.g., 3,000 energy levels × DOS values of 4-byte floats in our DFT calculations), it is very challenging to map the DOS pattern with only common material features as an input information, such as the number of atoms, composition, and lattice parameter. Thus, it is necessary to reduce the DOS patterns to a low-dimensional vector. To do this, we applied PCA in this work with a process similar to that of Yeo et al.[30] Prior to the analysis, the training DOS data were digitalized with 0 or 1 in a rectangular window of $M \times N$ grids, where they were cropped in the energy range of -10 eV to 3 eV relative to the Fermi level (0 eV) and the DOS range of 0 to 10. We standardized the DOS image vectors of the training data by obtaining the normalized matrix $\mathbf{Y}$, in which the $i^{th}$ row ($\mathbf{y}_i$) of $\mathbf{Y}$ is $\mathbf{x}_i - \bar{\mathbf{x}}$, where $\bar{\mathbf{x}}$ is the mean of each column vector of $\mathbf{X}$. Then, we calculated the principal components (PCs) or eigenvectors, $\mathbf{u}_p = (u_1, u_2, \ldots, u_{M \times N})_p$, and the corresponding eigenvalues, $\lambda_p$, were calculated by the covariance matrix, $\mathbf{S} = \mathbf{Y}^T\mathbf{Y}$, and Eq. (1).

$$\mathbf{S}\mathbf{u}_p = \lambda_p \mathbf{u}_p \tag{1}$$

The original DOS image vector $\mathbf{x}$ can be reconstructed as follows:

$$\mathbf{x} \cong \sum_{p=1}^{P}(\mathbf{y}^T\mathbf{u}_p)\mathbf{u}_p + \sum_{p=1}^{P}\left(\bar{\mathbf{x}}^T\mathbf{u}_p\right)\mathbf{u}_p = \sum_{p=1}^{P}\alpha_p\mathbf{u}_p \tag{2}$$

where P is the number of used PCs and $p$ is their index. Thus, coefficient $\alpha_p$ of the eigenvectors can be computed by $\mathbf{y}^T\mathbf{u}_p + \bar{\mathbf{x}}^T\mathbf{u}_p$, corresponding to the coordinate values on the linear subspace that is composed of PCs. In other words, the signal vector $\boldsymbol{\alpha} = (\alpha_1, \alpha_2, \alpha_3, \cdots, \alpha_P)^T$ can be defined as a one-to-one correspondence vector of $\mathbf{x}$. More details on the learning process of DOS patterns using PCA can be found in Ref. 30. However, in our new scheme, we extracted the signal vectors for partial DOS patterns of each atom in NP structures by the PCA process, while Yeo et al.[30] considered total DOS patterns.

In predicting the DOS pattern of a test system, we used the CGCNN[33] model to determine the new signal vector for the test system (Fig. 1), unlike Yeo et al.[30] Following the original CGCNN scheme, graphs for NP structures were constructed with nodes and edges, in which the nodes and edges represented atoms and bonds, respectively. In the graph, the atom vector $\boldsymbol{v}_i$ was encoded in a one-hot manner only with features that were readily available from the periodic table of elements (e.g., period/group number, melting temperature, etc.) due to their categorical property. The bond vector $\boldsymbol{u}_{(i,j)}$ was also encoded in a one-hot manner based on the bond length between atoms, in which the bond between the i$^{th}$ and j$^{th}$ atoms was defined only if $d_{i,j} < r_i + r_j + \Delta$, where $d_{ij}$ is a distance between the atoms i and j, and $r_i$ and $r_j$ are the radii of atoms i and j, respectively, with the tolerance $\Delta = 0.25$ Å. A list of the input features for the atom and bond vectors and their ranges/categories is available in Supplementary Tables S1 and S2.

Then, CNN processes were performed on top of the constructed graph, which consisted of a sequence of convolutions. The convolution functions first concatenated neighbor vectors $z_{(i,j)}^t = v_i^t \oplus v_j^t \oplus u_{(i,j)}$ and then performed convolutions to update each atom vector, as follows:

$$v_i^{t+1} = v_i^t + \sum_j \sigma(z_{(i,j)}^t W_f^t + b_f^t) \odot g(z_{(i,j)}^t W_s^t + b_s^t) \qquad (3)$$

where *t* denotes the number of convolutional layers; $\oplus$ denotes concatenation; $\odot$ denotes element-wise multiplication; $\sigma$ is a sigmoid function; g is the rectified linear unit (ReLU) function; and $W_f^t$, $W_s^t$ and $b_f^t$, $b_s^t$ are convolutional weight metrics and biases of the $i^{th}$ layer, respectively. After the convolution, the atom vectors for each atom learned with surrounding atoms and bonds in the NP structures can be extracted. Then, the learned atomic vectors were fully connected with the atomic signal vectors obtained from PCA via a neural network, in which the processes were performed for each atom in the training NP systems. Then, the total DOS pattern for a given NP structure was reconstructed through a summation of the partial DOS patterns mapped by the PCA-CGCNN architecture. In addition, the optimization of the CGCNN hyperparameters is justified in the Supplementary Information.

To validate our PCA-CGCNN model, we start with pure metallic NPs (Au, Pd, and Pt). Fig. 2 shows various NP structures used in the training and test sets, where not only small NPs but also bulk and slab structures were considered to train various local atomic environments. For example, various crystal (bulk) structures of metal elements (e.g., face and body-centered cubic, simple cubic, and diamond) were considered to learn the effects of atomic coordination numbers on the DOS patterns, and the effects of bond distances on the DOS patterns were also learned by considering crystal structures strained from -4% to +4% of their lattice parameters.

A comparison of the DOS patterns of Au NPs obtained from the DFT method and the PCA-CGCNN model is shown in Fig. 3. For the Au NPs, the similarities ($R^2$ values) of the DOS patterns reconstructed from the PCA-CGCNN model are in the ranges of 0.82~0.99 for the training systems and 0.83~0.90 for the test systems (Fig. 3a). The similarity is overall increased as the NP size becomes larger, which can be understood from the fact that a larger NP has lower surface fraction. Because surface atoms in NPs have different chemical environments (e.g., coordination numbers and bond lengths) than core atoms, it is likely more challenging to map DOS patterns of the smaller NPs in a given dataset. Considering that the computation cost of the DFT calculation is significantly increased with the size of NPs,[29,34] the superior prediction capability of the PCA-CGCNN model for larger NPs becomes a strong advantage of this model in terms of computational efficiency. In Figs. 3b and c, the DOS patterns of $Au_{85}$ and $Au_{108}$ NPs are shown, where the DOS similarities of the PCA-CGCNN model are 0.99 and 0.90, respectively. Indeed, our ML scheme reasonably reproduces the DFT pattern; in particular, the peak positions are very well matched, although only a handful of training structures are considered. For pure Pt and Pd NPs, our ML scheme demonstrates similar predictive abilities to those observed in the Au NPs (Supplementary Figs. S3 and S4), which clearly validates our PCA-CGCNN method.

To examine the transferability of our PCA-CGCNN method to bimetallic systems, Pd-Pt and Pt-Au binary systems were explored, in which not only core@shell structures but also alloyed structures were considered. In training DOS patterns for the systems, we used training DBs including pure and alloyed systems (Supplementary Fig. S5). When learning DOS patterns in each bimetallic system, we first applied PCA for atoms in training systems together, hereafter called total ML. The ML model for the Au@Pt systems provides low DOS similarities. Even for $Au_6@Pt_{32}$ in the training set, the DOS similarity value is so low that it is only 0.23. (Fig. 4). For other core@shell-type systems, similar behaviors are observed (Supplementary Figs. S6-S8), although the alloyed Pd-Pt systems show fair DOS similarities (Supplementary Fig. S9).

To improve the prediction ability of our PCA-CGCNN model, we propose a separate learning scheme during the PCA algorithm. For example, when predicting the DOS patterns of Pt-Au NP systems, the original PCA-CGCNN model was simultaneously trained with the DOS patterns of Pt, Au, and bimetallic Pt-Au NPs in the training set, and then the DOS patterns were predicted or reconstructed by the single model. However, in the separate learning scheme, the DOS patterns are individually trained for each atom, i.e., one model is trained with the atomic DOS patterns of Pt atoms in pure Pt and Pt-Au NPs, and another model is trained with those of Au atoms in pure Au and Pt-Au NPs. In the prediction process, the patterns of Pt atoms in bimetallic NPs are mapped with the Pt DOS-trained model, and the patterns of Au atoms are mapped with the Au DOS-trained model. Then, the mapped partial DOS patterns are summed to obtain the total DOS of each NP. In Fig. 4, the prediction ability of the PCA-CGCNN model for the Au@Pt NPs is significantly improved by the separate learning scheme. The DOS similarities of $Au_6@Pt_{32}$, $Au_6@Pt_{38}$, and $Au_{32}@Pt_{76}$ are 0.23, 0.52, and 0.41 from the total ML scheme, respectively; however, the separate ML scheme leads to 0.80, 0.89, and 0.78, respectively (Fig. 4a). Moreover, the DOS peak positions mapped by the separate ML scheme are much better matched with the DFT peaks than those mapped by the total ML scheme (Figs. 4b and 4c). Similar improvements are also observed in other bimetallic NP cases for all of the core@shell and alloyed structures (Supplementary Figs. S6-S9).

As already mentioned, DFT calculations of NP structures require an extremely increasing computational cost as the NP size increases. Thus, a comparison of computation speeds between DFT and the PCA-CGCNN method is of great interest. With an example of Pt NPs, we benchmark the computational speeds of each method (Fig. 5). Here, the DFT calculations were performed on 20 cores of a 2.3 GHz central processing unit (CPU), while the PCA-CGCNN calculations were performed on a personal computer with a single GTX 2070 graphics processing unit (GPU). In Fig. 5, it is clear that the PCA-CGCNN method is extremely fast compared with the DFT method. For example, for $Pt_{116}$ and $Pt_{147}$ NPs, the DOS calculations via the DFT method take 430 and 570 hours, respectively, which are much longer times than those for the PCA-CGCNN method (158 seconds for $Pt_{116}$ and 159 seconds for $Pt_{147}$). Here, the computational times for the PCA-CGCNN method are measured as a sum of training and prediction times. The PCA-CGCNN method takes only ~160 seconds (training: ~150 seconds, prediction: <10 seconds) for mapping the DOS patterns of NPs irrespective of the sizes of NPs, which is similar to the times reported for the previous PCA-only model[30]. This indicates that the addition of the CGCNN into the PCA method does not sacrifice the computational cost at all; instead, the addition of the CGCNN provides a more flexible and accurate approach.

In conclusion, we have developed the ML model combining PCA and the CGCNN to predict the DOS patterns of various types of NPs with a handful of training sets that can be obtained without great difficulty by the typical DFT frameworks, in which the PCA-CGCNN method is applicable for not only pure NPs but also bimetallic NPs. Although there is a small loss of accuracy with our PCA-CGCNN method compared to DFT calculations, the prediction speed is much faster than those of typical DFT frameworks. In particular, the prediction speed is not nearly as affected by the system sizes of NPs when compared with DFT frameworks. In this regard, our ML approach can become an option to circumvent DFT calculations, with which one can predict the DOS patterns of actual nanometer-scale NPs mostly synthesized in experiments which remains challenging within DFT frameworks. Therefore, our approach can be immediately applied to accelerate material design in diverse nanotechnology fields such as catalysis, biomaterials, and optics. Moreover, because our approach provides a flexible framework in handling atomic structures, it can be generally used to explore the correlation between atomic structures and other spectrum-type properties of materials (e.g., X-ray diffraction, X-ray photoelectron spectroscopy, Raman spectroscopy, etc.).

**METHODS**

Density functional theory (DFT) calculations

To obtain electronic DOS patterns of crystal, slab and NP structures, spin-polarized DFT calculations with plane-wave basis sets were carried out using the Vienna Ab initio Simulation (VASP) package[35,36]. We used the generalized gradient approximation with the revised Perdew-Burke-Ernzerhof functional[37,38] to describe the exchange-correlation energy of electrons. Ionic cores were treated by the projector-augmented wave method[39]. The plane-wave cutoff was set to 520 eV, and the convergence criteria for electronic structure and atomic geometry were $1.0\times10^{-4}$ eV and 0.03 eV/Å, respectively. The effect of the core electrons was modeled by projector-augmented wave (PAW) potentials[39]. The Brillouin zone was sampled using a Monkhorst-Pack k-point mesh[40], and the $k$-point sampling was set to $12\times12\times12$ for bulk structures, $12\times12\times1$ for slab structures, and $1\times1\times1$ for NP structures. A large vacuum spacing >20 Å was used for NP and slab structures to prevent interslab interactions. The DOS patterns were normalized by the number of atoms in the system and were shifted to set the Fermi level ($E_f$) to 0 in the pattern.

Principal component analysis (PCA)

After the PCA process, the DOS patterns of NPs can be expressed with a simple model in a linear combination form of PC components. Although the available number of PCs in a learning process of DOS patterns is as many as the dimensions (200×200) of the DOS image vectors, only a small number of PCs highly contribute to the representation of the DOS, where the contribution can be evaluated with an eigenvalue for each PC. With 25 PCs, we can readily represent the diversity of the DOS patterns in the training data. For example, in the Au NPs of Fig. 3, the contribution of the 25 PCs to the representation of the DOS pattern is from 6.3% for the 1$^{st}$ PC and 5.8% for the 2$^{nd}$ PC to 0.9% for the 25$^{th}$ PC, leading to a total contribution of ~90%. Of course, the consideration of a higher number of PCs can increase the accuracy for mapping DOS patterns, although the calculation speed also increases.

The performance (calculation speed and accuracy) of our PCA method can be affected by not only the number of PCs but also the grid size. Thus, we optimized the grid size and then found that a 200×200 grid shows the lowest mean absolute error (MAE) for comparison of the DOS patterns of Pt NPs obtained from our ML methods and DFT calculations (Supplementary Fig. S1), which justifies the selection of the 200×200 grid in this study.

Hyperparameter optimization of the CGCNN

The hyperparameters of the CGCNN model were thoroughly tested. The hyperparameters with their optimized values shown in parentheses are as follows: the number of convolution filters and layers (1 filter, 3 layers), learning rate ($1\times10^{-3}$), exponentially decaying learning rate (0.97 for every 100 epochs), nodes of the hidden layers (2 layers with 30 nodes/layer), standard deviation of normally distributed random initial weights (0.005), batch size (64), and total number of epochs (1000). The loss function is set as the mean square error (MSE). Dropout[41] and $L^2$ regularization were applied to overcome overfitting, where the dropout property and $L^2$ regularization coefficients were 0.3 and 0.01, respectively.

For atom vectors, we considered the following features; group number, period number, atomic number, radius, electronegativity, ionization energy, electron affinity, volume, atomic weight, melting temperature, boiling temperature, density, $Z_{eff}$, polarizability, resistivity, capacity, the number of valence electrons, and the number of *d*-electrons. A list of the input features for the atom and their ranges/categories is available in Supplementary Table S1. To select appropriate features for the atom vector, we calculated the MAE for the signal vectors of Pt as a function of the number of features (Supplementary Figure S2). For the cost efficiency, the best feature set was fixed by increasing the number of features. For example, the lowest MAE for the use of one feature was found with a melting temperature; thus, melting temperature was always included in the feature sets of the subsequent tests. From this optimization, the lowest MAE was found when only three features (melting temperature, ionization energy, and number of d-electrons) were used. Therefore, in this work, we used the three features for representing the atom vectors of Pt, Au, and Pd in the CGCNN. For bond vectors, we categorized distances between two atoms in the range of 2.4 to 4.0 Å into 20 dimensions (Supplementary Table S2).

DOS pattern similarity calculation

The DOS pattern similarity of our PCA-CGCNN model was calculated through the coefficient of determination ($R^2$) of the DOS pattern, which is defined as follows:

$$R^2 = \frac{\sum_m (\rho(E_m) - \rho'(E_m))^2}{\sum_m (\rho(E_m) - \bar{\rho})^2}$$

where ρ and ρ´ are the DOS patterns calculated by DFT and predicted by our PCA-CGCNN model, respectively, and $\bar{\rho}$ is the mean of DOS patterns calculated by DFT.

## DATA AVAILABILITY

All data in this work are available from the corresponding authors on reasonable request.

## Acknowledgment

This work was supported by Samsung Research Funding & Incubation Center of Samsung Electronics under Project Number SRFC-MA1801-03. This work was supported by the National Research Foundation of Korea (NRF) grant funded by the Korea government (MSIT) (No. 2017R1E1A1A03071049).

## AUTHOR CONTRIBUTIONS

S.S.H. and H.M.L. conceived and designed the research. K.B., B.C.Y. and D.K. implemented the PCA-CGCNN model. K.B. contributed to DFT data generation. K.B., S.S.H., and H.M.L. conceived a separate-learning method. All authors analyzed the data. K.B., S.S.H., and H.M.L. wrote the manuscript with feedback from all authors. S.S.H. and H.M.L. managed the project.

## ADDITIONAL INFORMATION

**Supplementary information** accompanies the paper on the *npj Computational Materials* website.

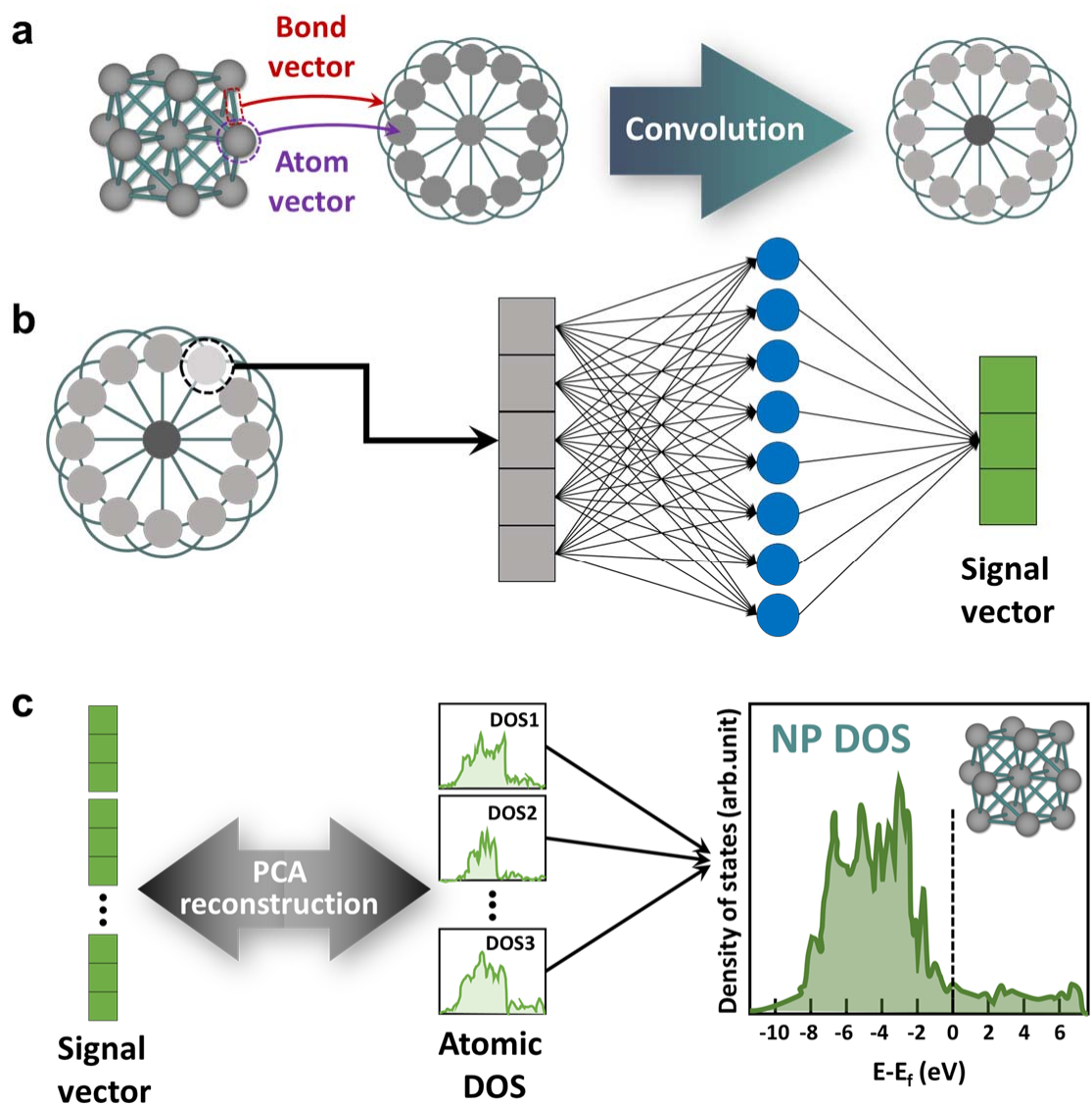

**Figure 1. Illustration of the PCA-CGCNN architecture**. **a** Construction of the crystal graph (CG) of NP structures and the structure of the convolutional neural network (CNN) on top of the CG. NP structures are converted to graphs with nodes and edges representing atoms and bonds, respectively. Then, the CNN processes are followed to reflect the local environments of each node in the CG. **b** Determination of signal vectors. After the CGCNN process, the new graph vector for each atom is fully connected with a signal vector for the DOS representation of each atom by neural networks. **c** DOS representation. With the signal vector obtained from the CGCNN, atomic DOS patterns are reconstructed on the basis of PCA. The sum of each atomic DOS pattern produces a total DOS pattern of the NP.

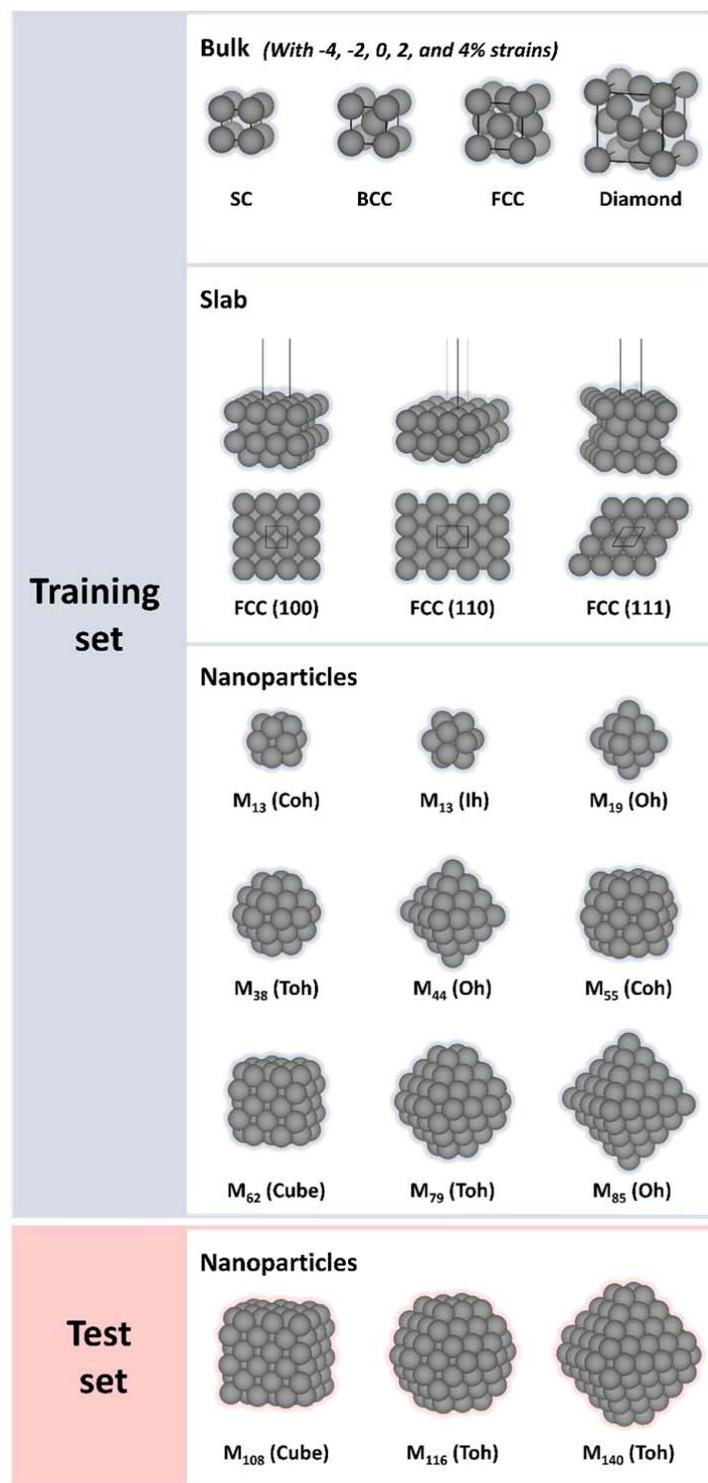

**Figure 2. Training and test datasets for DOS prediction of NPs.** In the NP structures, COh, Ih, Oh, TOh, and Cube indicate cuboctahedral, icosahedral, octahedral, tetraoctahedral, and cubic structures, respectively.

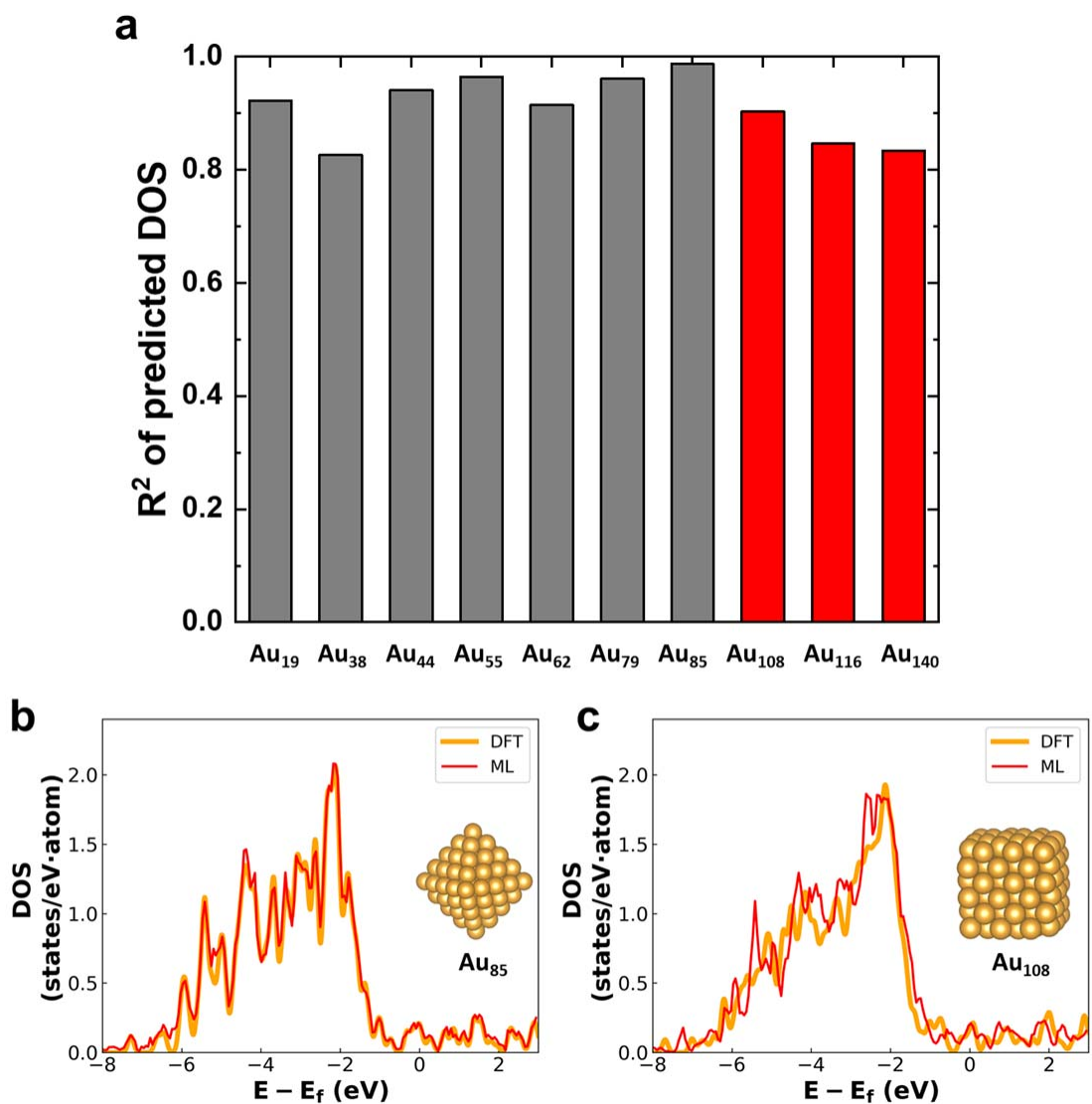

**Figure 3. PCA-CGCNN performance on Au NPs. a** The DOS pattern similarity ($R^2$ value) of our PCA-CGCNN model compared to DFT methods. Here, pure Au NPs are considered. Gray bars indicate training data, and red bars indicate test data. **b, c** Comparison of DOS patterns for **b** $Au_{85}$ and **c** $Au_{108}$ NPs predicted by the DFT method (yellow) and the PCA-CGCNN model (red).

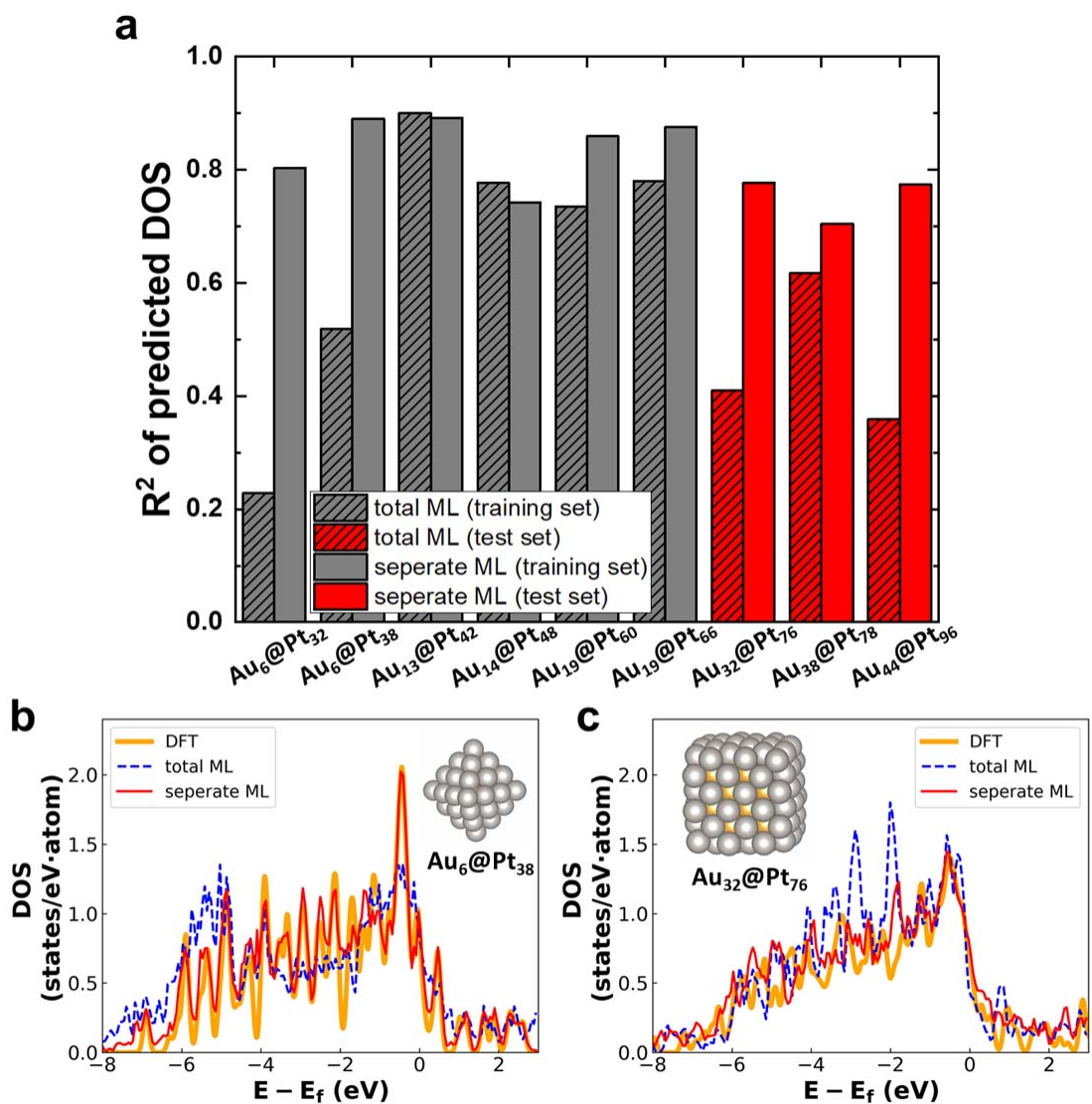

**Figure 4. PCA-CGCNN performance on Au@Pt bimetallic NPs. a** The DOS pattern similarity ($R^2$ value) of our PCA-CGCNN model compared to DFT methods. Here, bimetallic Au@Pt NPs are considered. **b, c** Comparison of DOS patterns for **b** $Au_6@Pt_{38}$ and **c** $Au_{32}@Pt_{76}$ NPs predicted by the DFT method (yellow) and the PCA-CGCNN model (blue = total learning and red = separate learning).

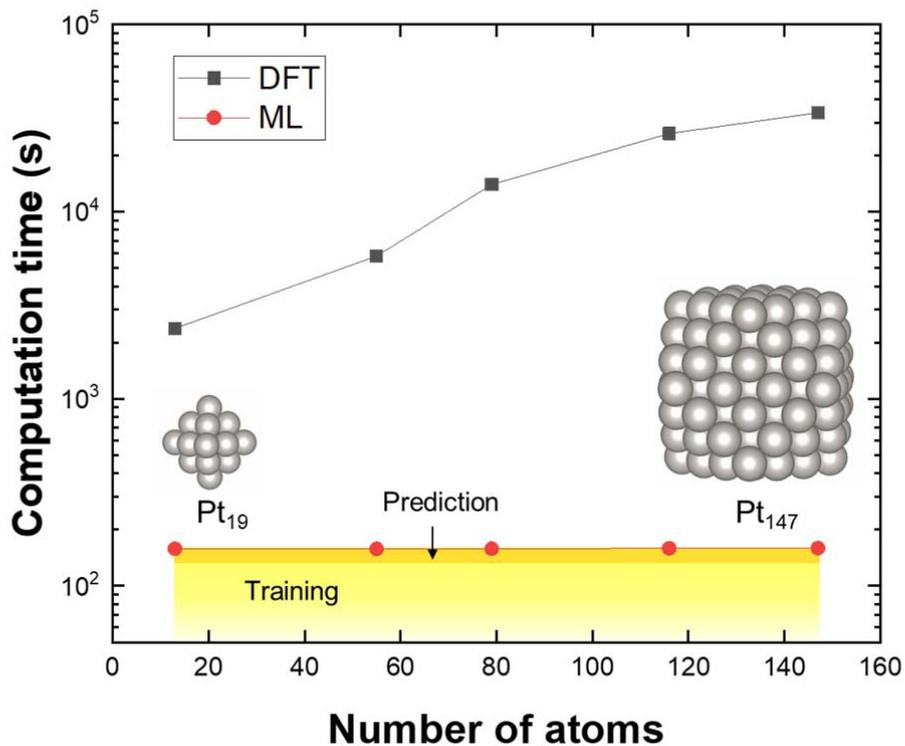

**Figure 5. Computational cost of PCA-CGCNN methods.** Comparison of the computation time for calculations of DOS patterns of metallic NPs via DFT (black) and ML (red). For the DFT calculations, a 2.3 GHz 20 core CPU was used. For the PCA-CGCNN methods, a personal computer with a GTX 2070 GPU and i5-9600K CPU was used, and the computational times were measured as a sum of training and prediction times.